\documentclass[12pt]{article}
\usepackage{epsf}
\def\beq{\begin{equation}}
\def\eeq{\end{equation}}
\def\bea{\begin{eqnarray}}
\def\eea{\end{eqnarray}}
\def\bq{\begin{quote}}
\def\eq{\end{quote}}

\def\bq{\begin{quote}}
\def\eq{\end{quote}}

\def\bq{\begin{quote}}
\def\eq{\end{quote}}

\def\gappeq{\mathrel{\rlap {\raise.5ex\hbox{$>$}}
{\lower.5ex\hbox{$\sim$}}}}

\def\lappeq{\mathrel{\rlap{\raise.5ex\hbox{$<$}}
{\lower.5ex\hbox{$\sim$}}}}

\def\bbz{fa Z \kern-8.9pt Z}

\begin{document}

\baselineskip 24pt
\newcommand{\sheptitle}
{Neutrino Masses in SUSY Models
\footnote{Invited Talk at EuroConference on Frontiers
in Particle Astrophysics and Cosmology, San Feliu de Guixols, Spain,
30 September - 5 October 2000}
}

\newcommand{\shepauthor}
{S. F. King }

\newcommand{\shepaddress}
{Department of Physics and Astronomy,
University of Southampton, Southampton, SO17 1BJ, U.K.}

\newcommand{\shepabstract}
{In this talk we assume the conventional see-saw mechanism,
and contruct a hierarchical pattern of three active neutrinos
with bi-maximal mixing. In order to enforce the hierarchy we
use the single right handed neutrino dominance
mechanism which is a very nice way to ensure
a neutrino mass hierarchy in the presence of large mixing angles.
We show how this mechanism can be organised in the framework
of a $U(1)$ family symmetry, then discuss a 
realistic string-inspired model which includes such a symmetry.}

\begin{titlepage}
\begin{flushright}
hep-ph/0011369\\
\end{flushright}
\begin{center}
{\large{\bf \sheptitle}}
\bigskip \\ \shepauthor \\ \mbox{} \\ {\it \shepaddress} \\ \vspace{.5in}
{\bf Abstract} \bigskip \end{center} \setcounter{page}{0}
\shepabstract
\end{titlepage}

\section{Introduction}

The problem of understanding the quark and lepton masses and
mixing angles represents one of the major unsolved questions
of the standard model. Recently additional information on
the fermion mass spectrum has come from the measurement of the
atmospheric neutrino
masses and mixing angles by Super-Kamiokande \cite{SKamiokandeColl}.
The most recent data disfavours mixing involving a sterile
neutrino, and finds a good fit for $\nu_{\mu} \rightarrow \nu_{\tau}$
mixing with $\sin^22\theta_{23}>0.88$ and a mass square splitting
$\Delta m^2_{23}$ in the $1.5-5\times 10^{-3} {\rm\ eV}^2$
range at 90\% CL \cite{HSobel}. 
Super-Kamiokande has also provided additional support for
solar neutrino mixing. The most recent Super-Kamiokande
data does not show a significant
day-night asymmetry and shows an energy independent
neutrino spectrum, 
thus it also disfavours the sterile neutrino mixing
hypothesis, the just-so vacuum oscillation hypothesis, and
the small mixing angle (SMA) MSW \cite{MSWMechanism}
solution \cite{YSuzuki}.
The preferred solution at the present time seems to be the
large mixing angle (LMA) MSW solution, although a similar
solution with a low mass splitting (the LOW) solution is also
possible. A typical point in the LMA MSW region is 
$\sin^22\theta_{12}\approx 0.75$, and $\Delta m^2_{12}\approx 2.5\times
10^{-5} {\rm\ eV}^2$ \cite{BaKrSm}.

If one accepts the recent data as evidence for neutrino masses and
mixing angles, then the obvious question is how these can be
accommodated in the standard model, or one of its supersymmetric 
extensions. The simplest possibility to account for the smallness
of the neutrino masses is the see-saw mechanism \cite{seesaw}
in which one introduces right-handed neutrinos which acquire 
very large Majorana masses at a super-heavy mass scale.
When one integrates out the right-handed neutrinos the ``normal sized''
Dirac Yukawa couplings, which connect the left-handed to the right-handed
neutrinos, are transformed into very small couplings
which generate very light effective left-handed physical 
Majorana neutrino masses.
Given the see-saw mechanism, it is natural to expect that the
spectrum of the neutrino masses will be hierarchical, since the 
Dirac Yukawa couplings in the charged fermion sector are observed
to be hierarchical, and if they are related to the Dirac neutrino
Yukawa couplings then they should also be hierarchical,
leading to hierarchical light Majorana masses.
\footnote{However this is not guaranteed due to the unknown
structure of the heavy Majorana matrix, and for example an inverted
neutrino mass hierarchy could result although this 
relies on some non-hierarchical couplings in the Dirac 
Yukawa matrix \cite{KiSi2}.}

Having assumed the see-saw mechanism and a hierarchical 
neutrino mass spectrum, the next question is how such large
(almost maximal) lepton mixing angles such as $\theta_{23}$ could
emerge? There are several possibilities that have been suggested
in the literature. One possibility is that it happens
as a result of the off-diagonal 23 entries in the
left-handed Majorana matrix being large, 
and the determinant of the
23 sub-matrix being accidentally small, leading to a neutrino
mass hierarchy with large neutrino mixing angles \cite{ElLeLoNa}.
Another possibility is that the neutrino mixing angles start
out small at some high energy scale, then get magnified 
by renormalization group (RG) running down to low energies
\cite{BaLePa:MTanimoto}. A third possibility is that the 
off-diagonal elements of the left-handed neutrino Majorana matrix
are large, but the 23 sub-determinant of the matrix is small
for a physical reason, as would be the case if a single right-handed
neutrino were providing the dominant contribution to the
23 sub-matrix \cite{KingSRND,SFKing1,SFKing2,KS}. 
We shall refer to these three approaches
as the accidental, the magnification and the single right-handed
neutrino dominance (SRHND) mechanisms, respectively.

A promising approach to understanding the fermion mass spectrum is within
the framework of supersymmetric (SUSY) unified theories.
Within the framework of such theories the quark and lepton masses
and mixing angles become related to each other, and it begins
to be possible to understand the spectrum. The simplest
grand unified theory (GUT) is $SU(5)$ but this theory in its
minimal version does not contain any right-handed neutrinos.
Nevertheless three right-handed neutrinos may be added, and in this theory
it is possible to have a large 23 element 
\footnote{We use Left-Right (LR) convention for Yukawa matrices
in this paper.} on the Dirac neutrino
Yukawa matrix without introducing a large 23 element into any of the
charged fermion Yukawa matrices. The problem of maintaining
a 23 neutrino mass hierarchy in these models may be solved for example by
assuming SRHND \cite{AlFe2}. Another possibility within the
framework of $SU(5)$ is to maintain all the off-diagonal
elements to be small, but require the 22 and 32 elements of
the Dirac neutrino Yukawa matrix to be equal and the
second right-handed neutrino to be dominant, in which case
SRHND again leads to a large 23 neutrino mixing angle with 
hierarchical neutrino masses \cite{AlFeMa}.
However the drawback of $SU(5)$ is that it does not predict
any right-handed neutrinos, which must be added as an afterthought.

From the point of view of neutrino masses, the most natural
GUTs are those like $SO(10)$ that naturally predict right-handed
neutrinos. However within the framework of $SO(10)$ the
quark masses and mixing angles are related to the lepton 
masses and mixing angles, and the existence
of large neutrino mixing angles is not expected in the minimal versions
of the theory in which the Higgs doublets are in one (or two) ${\bf 10}'s$
(ten dimensional representations of $SO(10)$) 
and each matter family is in a ${\bf 16}$.
Nevertheless various possibilities
have been proposed in $SO(10)$ in order to account for 
the large neutrino mixing angles. Within the framework of
minimal $SO(10)$ with third family Yukawa unification, 
it has been suggested that if two operators with different
Clebsch coefficients contribute with similar strength then,
with suitable choice of phases,
in the case of the lepton Yukawa matrices one may
have large numerical 23 elements, which add up to give a
large lepton mixing angle, while for the quarks the 23 elements can be
small due to approximate cancellation of the two contributing
operators \cite{BaPaWi}. This is an example of the accidental mechanism
mentioned above, where in addition one requires the quark mixing
angles to be small by accident, although it remains to be seen if
the LMA MSW solution could be understood in this framework.
Moving away from minimal $SO(10)$,
one may invoke a non-minimal Higgs sector in which one Higgs
doublet arises from a ${\bf 10}$ and one from a ${\bf 16}$, and in this 
framework it is possible to understand atmospheric neutrino
mixing \cite{AlBa}. 
Alternatively, one may invoke a non-minimal matter
sector in which parts of a quark and lepton family arise from
a ${\bf 16}$ and other parts from a ${\bf 10}$, and in these models 
one may account for atmospheric and solar neutrinos
via an inverted mass hierarchy mechanism \cite{ShTa}.

We have recently discussed \cite{KO}
neutrino masses and mixing angles in a particular string-inspired
{\em minimal} model based on the Pati-Salam \cite{PaSa}
$SU(4)\times SU(2)_L \times SU(2)_R$ (422) group. 
As in $SO(10)$ the presence of the gauged $SU(2)_R$ predicts the
existence of three right-handed neutrinos.
However, unlike $SO(10)$, there is no Higgs doublet-triplet
splitting problem since in the minimal model
both Higgs doublets are contained in a $(1,2,2)$ representation.
Moreover, since the left-handed quarks and leptons are in the
$(4,2,1)$ and the right-handed quarks and leptons in the $(4,1,2)$ 
representations,
the model also leads to third family
Yukawa unification as in minimal $SO(10)$.
Although the Pati-Salam gauge group is not unified at the
field theory level, it readily emerges from string constructions
either in the perturbative fermionic constructions \cite{AnLe},
or in the more recent type I string constructions \cite{ShTy},
unlike $SO(10)$ which typically requires large Higgs representations
which do not arise from the simplest string constructions.
The question of fermion masses and mixing angles in the
string-inspired Pati-Salam model has already been discussed
for the case of charged fermions \cite{SFking2,AlKi3},
and later for the case of neutrinos \cite{AlKi4}.
For the neutrino study \cite{AlKi4} it was assumed that the 
heavy Majorana neutrino mass matrix was proportional to the
unit matrix, and only small neutrino mixing angles were considered.
Later on a $U(1)_X$ family symmetry was added to the model,
in order to understand the horizontal hierarchies, although in this
case the neutrino spectrum was not analysed at all \cite{AlKiLeLo1}.

We discussed \cite{KO} neutrino masses
and mixing angles in the string-inspired Pati-Salam model
supplemented by a $U(1)_X$ flavour symmetry. 
The model involves third family Yukawa unification and predicts
the top mass and the ratio of the vacuum expectation values $\tan \beta$,
as we also recently discussed in Ref.~\cite{KiOl2}. It was already known that
the model can provide a successful description of
the CKM matrix and predicts the down and strange
quark masses, although our up-dated
analysis differed from that presented previously \cite{AlKiLeLo1}
partly due to the recent refinements in third family Yukawa unification 
\cite{KiOl2}, but mainly as a result of the recent Super-Kamiokande data
which has important implications for the flavour structure of the model.
In fact our main focus was on the neutrino masses
and mixing angles which were not previously discussed at all in this
framework. We assumed a minimal version of the model, and avoided the use 
of the accidental cancellation mechanism, which in any case has
difficulties in accounting for bi-maximal neutrino mixing.
We also showed that the mixing angle magnification mechanism
can only provide limited increases in the mixing angles, due
to the fact that the  unified third family Yukawa coupling
is only approximately equal to 0.7 \cite{KiOl2} and is therefore too
small to have a dramatic effect. Instead, we reled on the
SRHND mechanism, and we showed how this mechanism
may be implemented in the 422 model by appropriate use of 
operators with ``Clebsch zeros''
resulting in a natural explanation for atmospheric neutrinos
via a hierarchical mass spectrum. We specifically
focused on the LMA MSW solution since this is preferred by the
recent fits.

\section{Single Right-handed Neutrino Dominance (SRHND)}

Consider two RH neutrinos for simplicity. Then let us write the
(Dirac) Yukawa couplings in the LR basis as
\begin{equation}
Y_{\nu}=
\left( \begin{array}{ccc}
0 & a & {d}\\
0 & b & {e}\\
0 & c & {f}
\end{array}
\right)
\end{equation}
and the heavy Majorana mass matrix as
\begin{equation}
M_{RR}=
\left( \begin{array}{ccc}
0 & 0 & 0    \\
0 & X & 0 \\
0 & 0 & {Y}
\end{array}
\right) 
\end{equation}
Then using the see-saw formula for the light effective Majorana
mass matrix $m_{LL}=Y_{\nu}M_{RR}^{-1}Y_{\nu}^Tv_2^2$
where $v_2$ is the Higgs doublet vacuum expectation value,
\begin{equation}
m_{LL}
=
\left( \begin{array}{ccc}
{\frac{d^2}{Y}}+\frac{a^2}{X}
& {\frac{de}{Y}} +\frac{ab}{X}
& {\frac{df}{Y}}+\frac{ac}{X}    \\
.
& {\frac{e^2}{Y}} +\frac{b^2}{X}
& {\frac{ef}{Y}} +\frac{bc}{X}   \\
.
& .
& {\frac{f^2}{Y}} +\frac{c^2}{X}
\end{array}
\right){v_2^2}
\end{equation}

Now suppose that the third right-handed neutrino contributions 
$\sim { \frac{1}{Y}}$
dominate the 23 block of $m_{LL}$, then we get an automatic hierarchy
even for large 23 mixing:
\begin{equation}
det[m_{LL}]_{23}=m_{\nu_2}m_{\nu_3}\sim 0
\Rightarrow {m_{\nu_2}}/{m_{\nu_3}}\ll 1
\end{equation}
This physical mechanism responsible for the neutrino
mass hierarchy is called SRHND. In the limit that only a single right
handed neutrino contributes the determinant clearly exactly vanishes and we 
have $m_{\nu_2}=0$ exactly. In the actual situation the third
right-handed neutrino dominates, and sub-dominant contributions
from the second right-handed neutrino give a small finite
mass $m_{\nu_2}\neq 0$ (suitable for the MSW solution to the
solar neutrino problem) whilst maintaining the 23 mass hierarchy.
To be precise we have,
\begin{equation}
m_{\nu_1}=0,
\end{equation}
\begin{equation}
m_{\nu_2}\sim \frac{(b-c)^2}{X}v_2^2
\end{equation}
\begin{equation}
m_{\nu_3}\sim \frac{(d^2+e^2+f^2)}{Y}v_2^2
\end{equation}

The mixing angles may easily be estimated to be \cite{SFKing2}
\begin{equation}
\tan \theta_{23} \sim {\frac{e}{f}} \sim 1, 
\end{equation}
\begin{equation}
\tan \theta_{13} \sim {\frac{d}{\sqrt{e^2+f^2}}}\ll 1
\end{equation}
\begin{equation}
\tan \theta_{12} \sim { \frac{a}{b-c}} \sim 1  
\end{equation}
Thus by a suitable choice of Yukawa couplings it is possible to
have a large 12 angle suitable for the LMA MSW solution and a large
23 angle suitable for atmospheric oscillations, while maintaining
a small 13 angle consistent with the CHOOZ constraint.
Note that the bi-maximal mixing angle scenario does not
threaten the 23 mass hieararchy which is maintained by SRHND.
In other words, providing SRHND is in operation, the hierarchy
is guaranteed in the presence of large mixing angles.
This is a very nice feature of SRHND.

\section{$U(1)$ Family Symmetry}
Introducing a $U(1)$ family symmetry \cite{FN}, \cite{textures},
\cite{IR}, \cite{Ramond} provides a convenient way to
organise the hierarchies within the various Yukawa matrices.
For definiteness we shall focus on a particular class of model based
on a single pseudo-anomalous $U(1)$ gauged family symmetry \cite{IR}.
We assume that the $U(1)$ is broken by the equal VEVs of two
singlets $\theta , \bar{\theta}$ which have vector-like
charges $\pm 1$ \cite{IR}.
The $U(1)$ breaking scale is set by $<\theta >=<\bar{\theta} >$
where the VEVs arise from a Green-Schwartz mechanism \cite{GS} 
with computable Fayet-Illiopoulos $D$-term which
determines these VEVs to be one or two orders of magnitude
below $M_U$. Additional exotic vector matter with
mass $M_V$ allows the Wolfenstein parameter \cite{Wolf}
to be generated by the ratio \cite{IR}
\begin{equation}
\frac{<\theta >}{M_V}=\frac{<\bar{\theta} >}{M_V}= \lambda \approx 0.22
\label{expansion}
\end{equation}

The idea is that at tree-level the $U(1)$ family symmetry
only permits third family Yukawa couplings (e.g. the top quark
Yukawa coupling). Smaller Yukawa couplings are generated effectively
from higher dimension non-renormalisable operators corresponding
to insertions of $\theta$ and $\bar{\theta}$ fields and hence
to powers of the expansion parameter in Eq.\ref{expansion},
which we have identified with the Wolfenstein parameter.
The number of powers of the expansion parameter is controlled
by the $U(1)$ charge of the particular operator.
The fields relevant to neutrino masses
$L_i$, $N^c_p$, $H_u$, $\Sigma$
are assigned $U(1)$ charges $l_i$, $n_p$, $h_u=0$, 
$\sigma$. From Eqs.\ref{expansion},
the neutrino Yukawa couplings and Majorana mass
terms may then be expanded in powers of the Wolfenstein parameter,
\begin{equation}
M_{RR} \sim
\left( \begin{array}{ccc}
\lambda^{|2n_1+\sigma|} & \lambda^{|n_1+n_2+\sigma|} 
& \lambda^{|n_1+n_3+\sigma|}\\
. & \lambda^{|2n_2+\sigma|} & \lambda^{|n_2+n_3+\sigma|} \\
.  & .  & \lambda^{|2n_3+\sigma|} 
\end{array}
\right) 
\label{mRR}
\end{equation}
The conditions which ensure that the third dominant neutrino
is isolated require that the elements $\lambda^{|n_1+n_3+\sigma|}$,
$\lambda^{|n_2+n_3+\sigma|}$ be sufficiently small.

The neutrino Yukawa matrix is explicitly
\begin{equation}
Y_{\nu} \sim
\left( \begin{array}{ccc}
\lambda^{|l_1+n_1|} & \lambda^{|l_1+n_2|} 
& \lambda^{|l_1+n_3|}\\
\lambda^{|l_2+n_1|} & \lambda^{|l_2+n_2|} 
& \lambda^{|l_2+n_3|}\\
\lambda^{|l_3+n_1|} & \lambda^{|l_3+n_2|} 
& \lambda^{|l_3+n_3|}
\end{array}
\right)
\label{Ynu}
\end{equation}
The requirement
of large 23 mixing and small 13 mixing
becomes
\begin{equation}
|n_3+l_2|=|n_3+l_3|, \ \ \ \ |n_3+l_1|-|n_3+l_3|=1 \ or \ 2
\label{modulus}
\end{equation}

\section{Examples}
Let us consider an example of charges which lead to SRHND with
bimaximal mixing,
\begin{equation}
l_i=(-2,0,0),n_i=(-2,1,-1),\sigma=1
\end{equation}
then this leads to 
$$
m_{LL}
\sim
\left( \begin{array}{ccc}
\lambda^5+2\lambda^5
& \lambda^3 +\lambda^5+\lambda^3
& \lambda^3+\lambda^5+\lambda^3    \\
.
& \lambda +2\lambda^3
& \lambda +\lambda^3+\lambda^3  \\
.
& .
& \lambda +2\lambda^3
\end{array}
\right)
$$
where the first entry in each element corresponds to the $1/Y$ contributions
coming from the right-handed neutrino $N_{3R}$, and clearly
dominates the 23 block by a factor of $\lambda^2$.
In this case it does not dominate the other elements outside the 23 block.
The 13 element from $N_{3R}$ is suppressed relative to the 23 block elements
by a factor of $\lambda^2$, leading to a CHOOZ angle of this order.
The subdominant entries in the 12,13,22,23 elements are of the same
order, leading to a large 12 angle suitable for LMA MSW.

\section{Neutrino Masses and Mixing Angles
in a Realistic String-Inspired Model \cite{KO}}

Consider the Pati-Salam gauge group:
\begin{equation}
SU(4) \otimes SU(2)_L \otimes SU(2)_R\otimes U(1)
\end{equation}
with fermions in the superfield multiplets
\begin{equation}
{F^i_{L,R}}=
\left(\begin{array}{cccc} 
{u}  & {u} & {u} & {\nu} \\  
{d} & {d} & {d} & {e^-}     
\end{array} \right)_{L,R}^i     
\end{equation}

The model has the following features:
\begin{itemize}
\item{ predicts three right-handed neutrinos}
\item{ has third Family Yukawa Unification}
\item{ gives vertical hierarchies from Clebsch coefficients}
\item{ horizontal hierarchies from $U(1)$ family symmetry}
\item {this gauge group has emerged
from explicit type I string constructions.}
\end{itemize}

The Higgs $h$ contains the two MSSM Higgs doublets
\begin{equation}
h=
\left(\begin{array}{cc}
{h_1}^0 & {h_2}^+ \\   
{h_1}^- & {h_2}^0      
\end{array} \right) 
\end{equation}

The Higgs $H,\bar{H}$ break the Pati-Salam group
and $\theta, \bar{\theta}$ break $U(1)$ family symmetry.
\begin{equation}
{H},\bar{H} =
\left(\begin{array}{cccc}
{u_H} & { u_H} & { u_H} & { \nu_H} \\
{ d_H} & { d_H} & { d_H} & { e_H^-}   
\end{array} \right),\cdots
\end{equation}
\begin{equation}
<H>=<\bar{H}>=<{ \nu_H}>\sim M \sim 10^{16}GeV
\end{equation}
\begin{equation}
<\theta>=<\bar{\theta}>\sim M \sim 10^{16}GeV
\end{equation}
$$
SU(4)_{PS}\otimes SU(2)_L \otimes SU(2)_R \otimes U(1)
$$
$$
\longrightarrow SU(3)_C\otimes SU(2)_L \otimes U(1)_Y
$$

The fermion mass operators (responsible for Yukawa
matrices) are:
\begin{equation}
(F^i_L \bar{F}^j_R )h\left(\frac{H\bar{H}}{M^2}\right)^n
\left(\frac{\theta}{M}\right)^p
\end{equation}

The Majorana mass operators (responsible for $M_{RR}$) are:
\begin{equation}
(\bar{F}^i_R\bar{F}^j_R )\left(\frac{HH}{M^2}\right)
\left(\frac{H\bar{H}}{M^2}\right) ^m
\left(\frac{\theta}{M}\right) ^q
\end{equation}

The $U(1)$ family symmetry charges we assume are:
$$
F^{1}_L=1,\  F^2_L=0,\  F^3_L=0
$$
$$
\bar{F}^1_R = 4,\ \bar{F}^2_R = 2,\ \bar{F}^3_R =0 
$$

{
\begin{center}
\begin{tabular}{ccccc}
\multicolumn{5}{c}{Approximate Structure of the Matrices:} \cr
\noalign{\medskip}
%\noalign{\hrule height\rulerheight}
\noalign{\smallskip}
%\noalign{\hrule height\rulerheight}
\noalign{\medskip}
$
\lambda_u(M_X)$ & $\sim$ & $\left(\matrix{
\delta^3\epsilon^5 & \delta^2\epsilon^3 & \delta^2\epsilon\phantom{^3} \cr
\delta^3\epsilon^4 & \delta^2\epsilon^2 & \delta^3 \cr
\delta^3\epsilon^4 & \delta^2\epsilon^2 & 1 \cr} \right)$ 
\\
\noalign{\smallskip}
$
\lambda_d(M_X)$ & $\sim$ & $\left(\matrix{
\phantom{^3}\delta  \epsilon^5 &
\delta^2\epsilon^3 & 
\delta^2\epsilon\phantom{^3} \cr
\phantom{^3}\delta  \epsilon^4 &
\phantom{^3}\delta  \epsilon^2 & 
\delta^2  \cr
\phantom{^3}\delta  \epsilon^4 & 
\phantom{^3}\delta  \epsilon^2 &
1  \cr} \right)$ 
\\
\noalign{\smallskip}
$
\lambda_e(M_X)$ & $\sim$ & $\left(\matrix{
\phantom{^3}\delta  \epsilon^5 &
\phantom{^3}\delta  \epsilon^3 & 
\phantom{^3}\delta  \epsilon \phantom{^3} \cr
\phantom{^3}\delta  \epsilon^4 &
\phantom{^3}\delta  \epsilon^2 & 
\delta^2           \cr
\phantom{^3}\delta  \epsilon^4 &
\phantom{^3}\delta  \epsilon^2 &
1                  \cr} \right)$ 
\\
\noalign{\smallskip}
$
\lambda_\nu(M_X)$ & $\sim$ & $\left(\matrix{
\delta^3\epsilon^5 &
\phantom{^3}\delta\epsilon^3 & 
\phantom{^3}\delta\epsilon \phantom{^3} \cr
\delta^3  \epsilon^4 &
\delta^2  \epsilon^2 & 
\delta           \cr
\delta^3  \epsilon^4 &
\delta^2  \epsilon^2 &
1                  \cr}\right)$ 
\\
\noalign{\smallskip}
$
M_{RR}(M_X)$ & $\sim$ & $\left(\matrix{
\delta  \epsilon^8 &
\delta  \epsilon^6 &
\delta  \epsilon^4 \cr
\delta  \epsilon^6 & 
\delta  \epsilon^4 & 
\delta  \epsilon^2 \cr
\delta  \epsilon^4 &
\delta  \epsilon^2 & 1 \cr}\right)$ 
\\
\noalign{\medskip}
%\noalign{\hrule height\rulerheight}
\noalign{\smallskip}
%\noalign{\hrule height\rulerheight}
\end{tabular}
\end{center}}
{ where}
$\delta=\frac{<H><\bar{H}>}{M^2}$,
$\epsilon =\frac{<\theta >}{M}=\frac{<\bar{\theta} >}{M}.$

If we assume similar expansion parameters 
$\delta \sim \epsilon \sim \lambda \sim 0.22$ then
the above matrices lead to SRHND with the third right-handed neutrino
giving the dominant contribution the 23 block of $m_{LL}$.

The neutrino masses at low energies are:
$$
m_{\nu_1} = 4.84\times 10^{-8} {\rm\ eV},
$$
$$
m_{\nu_2} = 5.79\times 10^{-3} {\rm\ eV}, 
$$
$$
m_{\nu_3} = 5.39\times 10^{-2} {\rm\ eV}.               
$$

It is interesting to consider the renormalisation group
evolution of neutrino mixing angles at different
scales:

{ $Q = M_X\sim 3\times 10^{16} {\rm\ GeV}$}\\
$\sin^2(2\theta_{12}) = 0.828$, $\sin^2(2\theta_{23}) = 0.890$,
$\sin^2(2\theta_{13}) = \phantom{-}0.025$ \\

{ $Q = M_{\nu_3}\sim 3\times 10^{14} {\rm\ GeV}$}\\
$\sin^2(2\theta_{12}) = 0.832$, $\sin^2(2\theta_{23}) = 0.908$,
$\sin^2(2\theta_{13}) = \phantom{-}0.027$ \\

{ $Q = M_Z$}\\
$\sin^2(2\theta_{12}) = 0.853$,
$\sin^2(2\theta_{23}) = 0.943$,
$\sin^2(2\theta_{13}) = \phantom{-}0.027 $

\section{Conclusions }

We have assumed the conventional see-saw mechanism,
and contructed a hierarchical pattern of three active neutrinos
with bi-maximal mixing. In order to enforce the hierarchy we
used the SRHND mechanism. This is a very nice way to ensure
a neutrino mass hierarchy in the presence of large mixing angles.

We showed how this mechanism can be organised in the framework
of a $U(1)$ family symmetry, then went on to discuss a 
realistic string-inspired model which includes such a symmetry.
The main features of our approach may be summarised as follows:

\begin{itemize}
\item { SRHND is a natural and general mechanism for
yielding hierarchical
neutrino masses in the presence of large 23 mixing.}
\item{ SRHND underpins success of string-inspired SUSY Pati-Salam.}
\item{ ``Clebsch zero'' operators play key role}
\item{ SRHND models are stable under radiative corrections,
but atmospheric mixing can be increased by several per cent.}
\item {SRHND underpins several other models in the literature
\cite{AlFe2,AlFeMa}.}
\end{itemize}


\begin{thebibliography}{9}

\bibitem{SKamiokandeColl}
Y. Fukuda et al, Super-Kamiokande Collaboration,
Phys. Lett. {\bf B433}, 9 (1998);
ibid Phys. Lett. {\bf B436}, 33 (1998);
ibid Phys. Rev. Lett. {\bf 81}, 1562 (1998).

%------

\bibitem{HSobel}
H. Sobel, talk presented at the XIX International Conference on
Neutrino Physics and Astrophysics, Sudbury, Canada, June 16-21, 2000.

%---- Refs. for Mikheyev-Smirnov-Wolfenstein MSW mechanism --

\bibitem{MSWMechanism}
L. Wolfenstein, Phys. Rev. {\bf D17}, 2369 (1978);
ibid Phys. Rev. {\bf D20}, 2634 (1979); 
S. P. Mikheyev, A. Y. Smirnov, Yad. Fiz. {\bf 42}, 1441 (1985)
[Sov. J. Nucl. Phys. {\bf 42}, 913 (1985)];
Nuovo Cimento {\bf 9C}, 17 (1986).

%------

\bibitem{YSuzuki}
Y. Suzuki, talk presented at the XIX International Conference on
Neutrino Physics and Astrophysics, Sudbury, Canada, June 16-21, 2000.

%---- Ref data for LMA SMA solution.

\bibitem{BaKrSm}
J. N. Bahcall, P. I. Krastev, A. Y. Smirnov,
Phys. Rev. {\bf D60} 093001 (1999) {\tt hep-ph/9905220}.

%----- Ref for see-saw -----------------------------------------

\bibitem{seesaw}
M. Gell-Mann, P. Ramond, R. Slansky,
in Sanibel Talk, CALT-68-709, Feb. 1979 
and in {\it Supergravity} (North Holland, Amsterdam 1979);
T. Yanagida in 
{\it Proc. of the Workshop on Unified Theory and Baryon 
     Number of the Universe}, KEK, Japan (1979);
R. N. Mohapatra, G. Senjanovic,
Phys. Rev. Lett. {\bf 44}, 912 (1980).

%------

\bibitem{KiSi2}
% INVERTED HIERARCHY MODELS OF NEUTRINO MASSES.
S. F. King, N. N. Singh, {\tt hep-ph/0007243}.

%-----Accidental---

\bibitem{ElLeLoNa}
% NEUTRINO TEXTURES IN THE LIGHT OF SUPERKAMIOKANDE DATA AND A 
% REALISTIC STRING MODEL.
J. Ellis, G. K. Leontaris, S. Lola, D. V. Nanopoulos,
Eur. Phys. J. {\bf C9}, 389 (1999);

%----- Ref Running of \sin^2(2\theta_{23}) ---------------------

\bibitem{BaLePa:MTanimoto}
% RENORMALIZATION OF THE NEUTRINO MASS OPERATOR.
K. S. Babu, C. N. Leung, J. Pantaleone,
Phys. Lett. {\bf B319}, 191 (1993);
% RENORMALIZATION EFFECT ON LARGE NEUTRINO FLAVOR 
% MIXING IN THE MINIMAL SUPERSYMMETRIC STANDARD MODEL.
M. Tanimoto, Phys. Lett. {\bf B360}, 41 (1995). 

%----- Refs for SRND

\bibitem{KingSRND}
%
% ATMOSPHERIC AND SOLAR NEUTRINOS WITH A HEAVY SINGLET.
S. F. King, Phys. Lett. {\bf B439}, 350 (1998). 
%
% LARGE MIXING ANGLE MSW AND ATMOSPHERIC NEUTRINOS 
% FROM SINGLE RIGHTHANDED NEUTRINO DOMINANCE AND U(1) FAMILY SYMMETRY.


%------ (alpha,beta,gamma,delta)

\bibitem{SFKing1}
% ATMOSPHERIC AND SOLAR NEUTRINOS FROM SINGLE RIGHTHANDED
% NEUTRINO DOMINANCE AND U(1) FAMILY SYMMETRY.
S. F. King, Nucl. Phys. {\bf B562}, 57 (1999).


%------ (LMAMSW)

\bibitem{SFKing2}
S. F. King, Nucl. Phys. {\bf B576}, 85 (2000).

\bibitem{KS}
S. F. King and N. Singh, Nucl. Phys. {\bf B591}, 3 (2000).

%------

\bibitem{AlFe2}
% A SIMPLE GRAND UNIFICATION VIEW OF NEUTRINO MIXING AND FERMION MASS MATRICES.
G. Altarelli, F. Feruglio, Phys. Lett. {\bf B451}, 388 (1999). 

%------

\bibitem{AlFeMa} 
% LARGE NEUTRINO MIXING FROM SMALL QUARK AND LEPTON MIXINGS.
G. Altarelli, F. Feruglio, I. Masina,
Phys. Lett. {\bf B472}, 382 (2000).

%---- pati model of neutrinos

\bibitem{BaPaWi}
K. S. Babu, J. C. Pati, F. Wilczek, Nucl. Phys. {\bf B566}, 33 (2000). 

%------
\bibitem{AlBa}
% EXPLICIT SO(10) SUPERSYMMETRIC GRAND UNIFIED MODEL 
% FOR THE HIGGS AND YUKAWA SECTORS.
C. H. Albright, S. M. Barr, Phys. Rev. Lett. {\bf 85}, 244 (2000). 
 
%-------
\bibitem{ShTa}
% BIMAXIMAL NEUTRINO MIXINGS AND PROTON DECAY 
% IN SO(10) WITH ANOMALOUS FLAVOR U(1).
Q. Shafi, Z. Tavartkiladze, Phys. Lett. {\bf B487}, 145 (2000). 

\bibitem{KO} 
S.F.King and M.Oliveira, hep-ph/0009287.


%---- Refs. for Pati-Salam Model. --------------------

\bibitem{PaSa}
J. C. Pati, A. Salam, Phys. Rev. {\bf D10}, 275 (1974).


\bibitem{AnLe}
I. Antoniadis and G. K. Leontaris, Phys. Lett. {\bf B216}, 333 (1989);
I. Antoniadis and G. K. Leontaris, and J. Rizos, Phys. Lett. {\bf B245}, 161 (1990).

%-------

\bibitem{ShTy}
% TEV SCALE SUPERSTRING AND EXTRA DIMENSIONS.
G. Shiu, S. H. H. Tye, Phys. Rev. {\bf D58}, 106007 (1998). 

%----- King operators in the 422 model
\bibitem{SFking2}
% SU(4) X SU(2)-L X SU(2)-R AS A SURROGATE SUSY GUT.
S. F. King, Phys. Lett. {\bf B325}, 129 (1994).

\bibitem{AlKi3}
%FERMION MASSES IN A SUPERSYMMETRIC SU(4) X SU(2)-L X SU(2)-R MODEL. 
B. C. Allanach, S.F. King, Nucl. Phys. {\bf B456}, 57 (1995). 

\bibitem{AlKi4}
% NEUTRINO MASSES AND MIXING ANGLES IN A 
% SUPERSYMMETRIC SU(4) X SU(2)-L X SU(2)-R MODEL.
B. C. Allanach, S. F. King, Nucl. Phys. {\bf B459}, 75 (1996).

%-----

\bibitem{AlKiLeLo1}
% YUKAWA TEXTURES IN STRING UNIFIED MODELS WITH SU(4) X O(4) SYMMETRY. 
B. C. Allanach, S. F. King, G. K. Leontaris, S. Lola,
Phys. Rev. {\bf D56}, 2632 (1997). 


%---- Ref Yukawa Unification in the 422 model

\bibitem{KiOl2}
S. F. King, M. Oliveira, {\tt hep-ph/0008183}.

\bibitem{FN}  C. D. Froggatt and H. B. Nielsen, Nucl. Phys. {\ B147} (1979)
277.

\bibitem{textures}  
H. Fritzsch, Phys. Lett. {\ 70B} (1977) 436; {\ B73}
(1978) 317; 
J. Harvey, P. Ramond and D.
Reiss, Phys. Lett. {\ B92} (1980) 309; 
C. Wetterich, Nucl. Phys. {\ B261}
(1985) 461; P. Kaus and S. Meshkov, Mod. Phys. Lett. {\ A3} (1988) 1251.

\bibitem{IR}  L. Ibanez and G.G. Ross, Phys. Lett. B332 (1994) 100;
P. Binetruy and P. Ramond, Phys. Lett. B350 (1995) 49, hep-ph/9412385.

\bibitem{Ramond}
J. Elwood, N. Irges and P. Ramond,
Phys. Rev. Lett. {\bf 81} (1998) 5064, hep-ph/9807228;
N. Irges, S. Lavignac and P. Ramond,
Phys. Rev. {\bf D58} (1998) 035003, hep-ph/9802334.


\bibitem{GS}
M. Green and J. Schwartz, Phys. Lett. B149 (1984) 117.

\bibitem{Wolf}
L. Wolfenstein, Phys. Rev. Lett {\bf 51} (1983) 1945.




\end{thebibliography}
\end{document}